# Multi-wavelength source at ITU-T grid based on ultra-flattened dispersion photonic crystal fibers

Arismar Cerqueira S. Jr.[1,*], J. D. Marconi[1], L. H. Gabrielli[1], A. A. Rieznik[1], H.E. Hernandez-Figueroa[1], H. L. Fragnito[1] and J. C. Knight[2]

[1]*Optics and Photonics Research Center, Unicamp, 13083-970, Campinas-SP, Brazil*
[2]*Centre for Photonics and Photonics Materials, Department of Physics, University of Bath, Claverton Down, Bath, BA2 7AY, UK.*
[*]*Corresponding author: arismar@dmo.fee.unicamp.br*

We report on the development of a multi-wavelength source locked to the ITU-T grid at 100 GHz spacing and based on ultra-flattened dispersion photonic crystal fibers. The ITU-T grid frequencies are generated by taking advantage of multiple four-wave mixing processes. We experimentally obtained 250 products spanning over 200nm by launching 03 strong pump signals. Simulations based on Split-Step Fourier Method have been carried out in order to investigate and improve the performance of this nonlinear optical device.

*OCIS codes*: 230.4320, 190.4380, 060.5295.

## *Introduction*

Optical fibers are very attractive as nonlinear medium because light is easily confined in small volumes with long interaction length. There are a lot of devices based on optical nonlinear effects, such as all optical wavelength converters [1], parametric amplifiers [2] and optical limiters [3]. These devices are based on the nonlinear mechanism of Four-Wave Mixing (FWM), in which four different waves interact in such way that their energy and momentum are conserved. The momentum conservation is often referred to as phase matching and depends strongly on the chromatic dispersion of the nonlinear medium, in our case the fiber. As the waves propagate through the fiber, FWM processes may occur involving the new generated waves, creating in this manner photons at further new frequencies. This is referred to as cascaded or multiple FWM [4]. For the simultaneous propagation of multiple intense pulses, FWM can provide an efficient mechanism for broadband redistribution of the energy to new wavelengths. This frequency cascading is formed by signals with well-defined frequency and phase differences.

Multiple four-wave mixing has been studied both theoretically and experimentally by many authors. From the theoretical point of view, Thompson et al. used a six waves model to describe the interaction of two pump waves and four sidebands generated by FWM processes [5]. Recently, McKinstrie et al have obtained approximated formulas to predict the amplitude of light waves involved in four-wave mixing cascades for the case of dispersion-less fibers when the waves are located near the zero dispersion wavelength (ZDW) of the fiber [6].

On the practical side, Fatome et al. proposed the use of kilometer lengths of optical fiber for the generation of high repetition rate pulse sources based on multiple FWM [7]. They reported a multiple four-wave mixing bandwidth of 90 nm. Frequency comb generator for optical frequency metrology [8] is another application of this nonlinear phenomenon.

We have recently investigated a highly efficient technique to generate broadband cascade FWM products by using not only traditional highly nonlinear fibers, but also photonic crystal fibers [4, 9]. We have demonstrated that by shortening the fiber length to a few meters and by employing a multi-section arrangement is possible to generate cascaded four-wave mixing products spanning over 300 nm. It has been obtained by injecting two strong pump waves, spaced by 6.3 or 2.5 nm, into very short optical fibers (some meters) with the pump close to the ZDW ($\lambda_{ZD}$). The advantages of using very short optical fibers result from three effects: reducing the relative phase difference $\Delta\Phi$ of non-polarization maintaining fibers; avoiding undesirable variation of the zero-dispersion wavelength and preserving the phase matching condition by keeping the product $\Delta\beta L$ small [4]. Furthermore, we have shown that the use of fibers with low dispersion slope can improve the conversion efficiency of the multiple FWM products [4].

In the current work we apply ultra-flattened dispersion photonic crystal fibers [10] to the development of multi-wavelength sources locked to ITU-T grid at 100 GHz spacing. Instead of using two pumps as in our previous works, we propose to use three pumps located at ITU-T grid in order to obtain high efficiency and large bandwidth of multiple FWM spaced by only 100 GHz. Experimental results and simulations based on a Split Step Fourier Method (SSFM) [11] demonstrate the improvement in the efficiency of multiple FWM obtained by using this new scheme of three initial pumps at ITU-T grid.

## Analysis of multiple four-wave mixing efficiency

We have carried out simulations based on SSFM for evaluating the efficiency of multiple FWM processes. Our performance parameter was the number of FWM products with an Optical Signal-to-Noise Ratio (OSNR) above 30 dB. For these initial simulations we have used just one fiber with the following data: length $L = 3$ m, ZDW $\lambda_{ZD} = 1560$ nm, nonlinear coefficient $\gamma = 10$ W$^{-1}$km$^{-1}$ and dispersion slope $S_0 = 0.006$ ps/nm$^2$km. We have used a resolution of 5 pm in all the simulations. Three different configurations have been analyzed as a function of the total input power:

- Configuration 1: two pump lasers at $\lambda_1 = 1559.6$ nm and $\lambda_2 = 1560.4$ nm with OSNR of 96 dB @ 12.5 GHz
- Configuration 2: three pumps $\lambda_1 = 1559.6$ nm, $\lambda_2 = 1560.4$ nm and $\lambda_3 = 1561.2$ nm with OSNR of 96 dB @ 12.5 GHz
- Configuration 3: three pumps $\lambda_1 = 1559.6$ nm, $\lambda_2 = 1560.4$ nm and $\lambda_3 = 1562.0$ nm with OSNR of 96 dB @ 12.5 GHz.
.

As shown in Fig. 1, for the same level of total input power the efficiency of multiple FWM process is significantly enhanced by using three pump waves in place of two pumps as in our previous works. Moreover, it is clear that this efficiency can be further improved by using the configuration 3, in which we apply different separations between the pump waves: 100 GHz between $\lambda_1$ and $\lambda_2$ and 200 GHz between $\lambda_2$ and $\lambda_3$. For example, for total power $P = 250$ W, the number of FWM products with OSNR above 30 dB is 24, 45 and 71 for the configurations 1, 2 and 3, respectively. By using the configuration 3 we ensure FWM products will continue spaced by 100 GHz with the advantage of obtaining an improved distribution of energy between the FWM products.

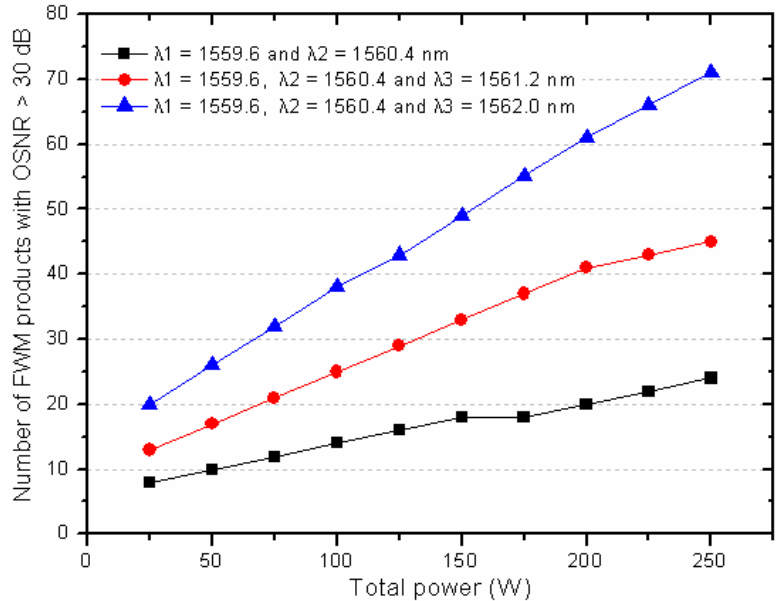

Fig. 1 Analysis of multiple FWM efficiency at ITU-T grid.

## *Multi-wavelength source at ITU-T grid*

We propose a multi-wavelength source at ITU-T grid based on the experimental setup shown in Fig.2. The light sources were three external cavity lasers (ECL) with a nominal linewidth of 120 kHz, coupled with a 33% coupler. The ECLs were tuned at $\lambda_1$ = 1561.01 nm, $\lambda_2$ = 1561.83 nm and $\lambda_3$ = 1563.45 nm. The lasers were modulated using a Mach Zehnder amplitude modulator (AM), obtaining 100 ns pulses at a repetition rate of 20 µs. Two erbium doped fiber amplifiers (EDFA) in cascade have been used to generate pulses of ~ 48.2 dBm peak power for each laser. The EDFA 1 is a low noise optical amplifier with an average output power up to 15 mW; whereas the EDFA 2 is a booster erbium amplifier with up to 1 W average output power. A polarization controller (PC) has been used to guarantee the parallelism among the polarization of the pumps, thus optimizing the efficiency of the four-wave mixing process. Light was finally coupled into the fiber by using an objective lens. The fiber was an ultra-flattened dispersion photonic crystal fiber (PCF), like those reported in [10], with 5 m length. Its measured zero dispersion wavelength, dispersion slope, attenuation and nonlinear coefficients are $\lambda_{ZD}$ = 1590 nm , $S_0$ = 0.006 ps/nm$^2$/km, $\alpha$ = 35 dB/km and $\gamma$ = 2 W$^{-1}$ km$^{-1}$, respectively. Output spectra were measured using an optical spectrum analyzer (OSA) with 0.01 nm resolution. The peak powers were monitored using a fast photodiode and an oscilloscope (OSC) with rise-time response < 1 ns.

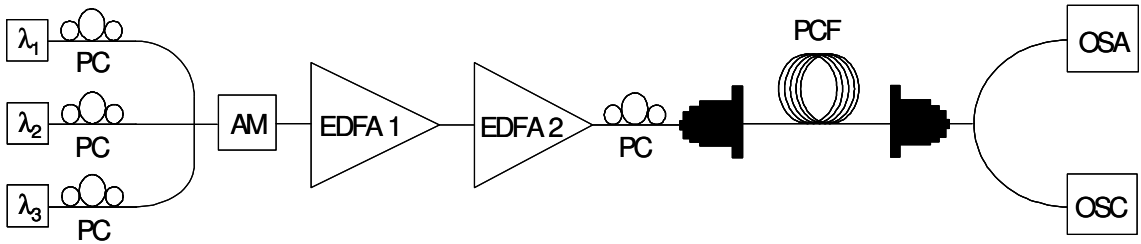

Fig. 2 Experimental setup.

We have previously reported that the first FWM peaks are already produced inside the EDFA booster [4]. This is important for the overall generation of multiple FWM products, since more than three waves will be launched into the fiber. In order to accurately analyze our multi-wavelength source, the measured spectrum at the booster output has been used as input spectrum of the ultra-flattened dispersion photonic crystal fiber in our simulations. By using the measured fiber parameters given above, we obtained the multiple FWM products shown in Fig. 3(a). This prediction is in good qualitative agreement with the experimental result as shown in Figs. 3(b) and 3(c), where we obtained 250 FWM products, extend from 1460 nm to 1660 nm and spaced by 0.8 nm (100 GHz) at the wavelengths of ITU-T grid specification. Fig. 3(c) shows a zoom at the spectral region correspondent to the C and L bands.

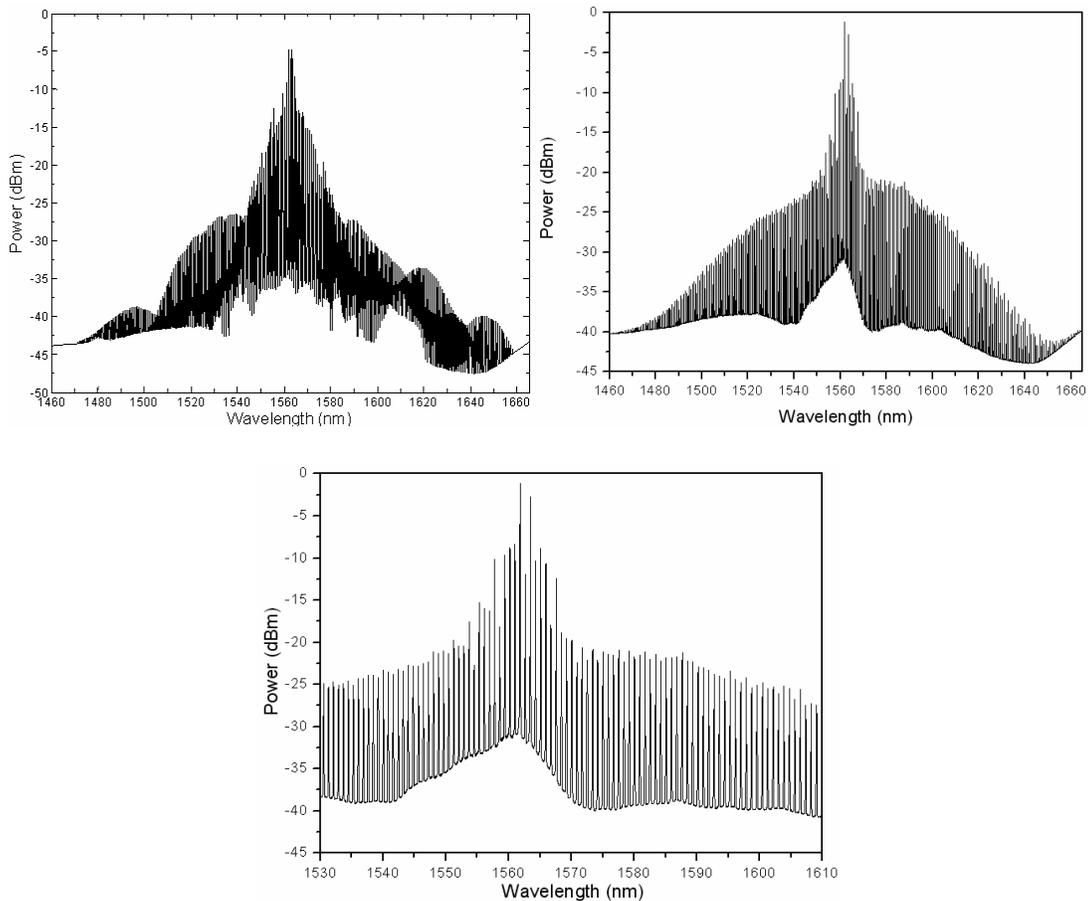

Fig. 3. Multi-wavelength source at ITU-T grid : (a) Simulated spectrum; (b) Experimental spectrum; (c) Zoom at the C and L-bands.

## Conclusions

A multi-wavelength source locked to ITU-T grid at 100 GHz spacing has been developed. It is based on broadband cascaded FWM products obtained by using ultra-flattened dispersion photonic crystal fibers. Experimental and simulated results show that the use of three properly spaced pumps in conjunction with fibers with a very low dispersion slope allowed us to work with the ITU-T spacing of 100 GHz. We have obtained 250 FWM products spanning over 200 nm, from 1460 to 1660nm. The FWM peaks are tuned at the frequencies of the ITU-T grid, covering entirely the C and L bands. This multi-wavelength light source can be efficiently used for testing a wavelength division multiplexing system.

## Acknowledgements

Authors acknowledge the help of Will Reeves at the University of Bath with fiber fabrication. The authors from Brazil thank the Brazilian agency FAPESP (Fundação de Amparo à Pesquisa e ao Ensino do Estado de São Paulo) for the financial support. Work at Bath was supported by the UK Engineering and Physical Sciences Research Council.